\documentclass[aps,superscriptaddress,prl,twocolumn]{revtex4-1}

\usepackage{mathptmx}
\usepackage{subfigure}
\usepackage{psfrag,graphicx}
\usepackage{mathptmx}
\usepackage{subfigure}
\usepackage{psfrag,graphicx}
\usepackage{dcolumn}
\usepackage{amsmath,amssymb}
\usepackage{bm}
\usepackage{color}
\usepackage{latexsym}
\usepackage{epstopdf}
\usepackage{color}
\usepackage[english]{babel}
\usepackage{latexsym}
\usepackage{multirow}

\usepackage{mathptmx}
\usepackage{subfigure}
\usepackage{psfrag,graphicx}
\usepackage{dcolumn}
\usepackage{amsmath,amssymb}
\usepackage{bm}
\usepackage{color}
\usepackage{latexsym}
\usepackage{epstopdf}
\usepackage{color}
\usepackage[english]{babel}
\usepackage{latexsym}
\usepackage{multirow}

\usepackage{psfrag,graphicx} 
\usepackage{epsf} 
\usepackage{subfigure} 
\usepackage{amsmath} 
\usepackage{amssymb} 
\usepackage{amsfonts}
\usepackage{bm}
\usepackage{natbib}
\usepackage{epstopdf}
\usepackage{appendix}

\usepackage{psfrag,graphicx} 
\usepackage{epsf} 
\usepackage{subfigure} 
\usepackage{amsmath} 
\usepackage{amssymb} 
\usepackage{amsfonts}
\usepackage{bm}
\usepackage{natbib}
\usepackage{epstopdf}
\usepackage{appendix}

\definecolor{mygrey}{gray}{0.35}
\definecolor{myblue}{rgb}{0.2,0.2,0.8}
\definecolor{myzard}{cmyk}{0,0,0.05,0}
\definecolor{mywhite}{rgb}{1,1,1}
\definecolor{myred}{rgb}{1,0.,0.3}

\usepackage[colorlinks=true,citecolor=myblue,linkcolor=myred]{hyperref}

\usepackage{geometry}
\usepackage{lineno}%[switch]
\makeatletter
\define@key{Gin}{Trim}
{\let\Gin@viewport@code\Gin@trim\expandafter\Gread@parse@vp#1 \\}
\makeatother
\newtoks\trimVal
\newtoks\trimVals
\def\be{\begin{equation}}
\def\ee{\end{equation}}
\def\ba{\begin{align}}
\def\enda{\end{align}}
\def\bi{\begin{itemize}}
\def\ei{\end{itemize}}

 \def\ee{\mathord{\rm e}}

 \def\ee{\mathord{\rm e}}

\renewcommand{\ee}{{\rm e}}

\def\beq{\begin{equation}}
\def\beq{\begin{equation}}
\def\eeq{\end{equation}}

 \newcommand{\ket}[1]{|#1\rangle}
 \newcommand{\bra}[1]{\langle #1|}

 \newcommand{\bea}{\begin{eqnarray}}
\newcommand{\eea}{\end{eqnarray}}

\usepackage{color}

\begin{document}

\title[Short Title]{Ultrasensitive Magnetometer Using a Single Atom}

\author{I. Baumgart}
\affiliation{Department Physik, Naturwissenschaftlich-Technische Fakult\"{a}t,  Universit\"{a}t Siegen, 57068 Siegen, Germany}
\author{J.-M. Cai}
\affiliation{School of Physics, Huazhong University of Science and Technology, Wuhan 430074, China}
\author{A. Retzker}
\affiliation{Racah Institute of Physics, The Hebrew University of Jerusalem, Jerusalem 91904, Givat Ram, Israel}
\author{M. B. Plenio}
\affiliation{Institut f{\"u}r Theoretische Physik, Universit{\"a}t Ulm, 89069 Ulm, Germany}
\author{Ch. Wunderlich*}
\affiliation{Department Physik, Naturwissenschaftlich-Technische Fakult\"{a}t,  Universit\"{a}t Siegen, 57068 Siegen, Germany}
\date{\today}

\begin{abstract}
Precision sensing, and in particular high precision magnetometry, is a central goal of research into quantum technologies. For magnetometers, often trade-offs exist between sensitivity, spatial resolution, and frequency range. The  precision, and thus the sensitivity of magnetometry, scales as $1/\sqrt {T_2}$ with the phase coherence time, $T_2$, of the sensing system playing the role of  a key determinant. Adapting a dynamical decoupling scheme that allows for extending $T_2$ by orders of magnitude and merging it with a magnetic sensing protocol, we achieve a measurement sensitivity even for high frequency fields close to the standard quantum limit. Using a single atomic ion as a sensor, we experimentally attain a sensitivity of $4.6$ pT $/\sqrt{\mbox{Hz}}$ for an alternating-current magnetic field near 14 MHz. Based on the principle demonstrated here, this unprecedented sensitivity combined with spatial resolution in the nanometer range and tunability from direct-current to the gigahertz range could be used for magnetic imaging in as of yet inaccessible parameter regimes. 

\end{abstract}

\maketitle

{\em Introduction --}
High precision measurements often have played a pivotal role for new discoveries in physics. Today, detecting electromagnetic fields with extreme sensitivity and spatial resolution is particularly important in condensed matter physics and in biochemical sciences. 
State-of-the-art magnetometers reach their best sensitivity in a limited frequency band or do not work at all (for all practical purposes) outside a certain frequency range. They often require a cryogenic and/or a carefully shielded environment. Also, their limited spatial resolution often makes them unsuitable for the applications mentioned above. Here, we introduce and demonstrate a novel method for sensing magnetic fields at the standard quantum limit, based on the use of a single atom as a sensor that is confined to a nanometer-sized region in space. The sensor can be tuned to a desired frequency where a signal shall be measured and is not affected by magnetic disturbances. Also, the magnetometer is essentially immune against amplitude fluctuations of the microwave fields that decouple the sensor from environmental disturbances.

Before introducing this novel magnetometer scheme and describing the experimental procedure, we briefly outline state-of-the-art magnetometry by means of a few examples.  Magnetic field sensitivities in the range of femto- or even subfemtotesla Hz$^{-1/2}$ have been reached using superconducting quantum interference devices \cite{Jaklevic1964}  or
atomic magnetometers \cite{Wasilewski2010,Kominis2003}. 
Optical atomic magnetometers  \cite{Cohen-Tannoudji1969,Budker2007} 
are alternative sensors based on the magneto-optical properties of atomic samples in vapor cells reaching a sensitivity in the fT  Hz$^{-1/2}$ range \cite{Weis2005}.
A persistent current quantum bit held at 43 mK was used to obtain a sensitivity of $3.3$ pT Hz$^{-1/2}$ measuring an ac magnetic field near 10 MHz \cite{Bal2012}. Often, detecting magnetic fields with the highest sensitivity and spatial resolution is mutually exclusive \cite{Wildermuth2005}. 
A high spatial resolution in the nanometer range with relatively low sensitivity is possible using sensors based on magnetic force microscopy \cite{Rugar1992}, 
or with Hall sensors \cite{Chang1992}. 
Using nitrogen vacancy centres in diamond,  
 nT Hz$^{-1/2}$ field sensitivity can be combined with (sub)nanometer spatial resolution \cite{Balasubramanian1,Grinolds2014,Pham2012}. 
Surface imaging with Bose-Einstein condensates reaches sensitivities of $\sim10$ pT Hz$^{-1/2}$ and $50$ $\mu$m spatial 
resolution \cite{Koschorreck2011}.

State-of-the-art magnetometry often relies on dynamical decoupling where fast pulses or continuous fields drive a quantum mechanical two-level system. The role of these fields is to decouple the system from the environment, and thus to enhance the $T_2$ time  \cite{Lloyd,Huelga1997}, while at the same time retaining the ability to sense a signal that is on resonance with the pulse rate or the Rabi frequency of the decoupling field. Random ambient magnetic field fluctuations, which are not featureless white noise but tend to have a limited bandwidth, are the dominant noise source in many cases and limit $T_2$. Pulsed dynamical decoupling (DD) was proposed and demonstrated for prolonging coherence times by subjecting a two-level system to a rapid succession of pulses leading to decoupling from the environment. This technique, often termed bang-bang control, originates in nuclear magnetic resonance experiments and can be applied in diverse systems \cite{Hahn1950,Viola1998,Biercuk2009,Hall2010,Kotler2011}. 

Throughout a dynamical decoupling pulse sequence the quantum probe is decoupled from ambient magnetic noise while increasing its sensitivity to alternating magnetic signals at specific frequencies. Thus, DD can be used to extract information about the magnetic noise spectrum \cite{Biercuk2009,Hall2010}, 
and to improve the signal-to-noise ratio in magnetic sensing by several orders of magnitude. Measuring high frequency components with high sensitivity  requires a high pulse rate and, in turn, shorter pulses with increased peak amplitude \cite{Balasubramanian1,Pham2012,Hall2010}. 
Using such a technique, a magnetometer sensitivity of 15 pT Hz$^{-1/2}$ for magnetic field frequencies up to 312.5 Hz was achieved using a single trapped ion as a probe \cite{Kotler2011} while with about $10^3$ nitrogen vacancy centers in diamond frequencies up to 220 kHz were measured with a sensitivity of order 10 nT Hz$^{-1/2}$ \cite{Pham2012}. 

Dynamical decoupling can also be achieved, in the so-called spin locking regime, via a simple continuous drive dressing atomic energy levels \cite{Goldman70,Rabl2009,lidar1}. 
This usually requires stabilization methods to decrease the effect of amplitude noise in the dressing field on the sensitivity \cite{Cai1}.
However, it has been recently demonstrated using trapped ions \cite{Timoney2011,Webster2013,Randall2015,Mikelson2015}
that by using  additional atomic levels the effect of  noise in the dressing fields can be dramatically reduced, and extensions were proposed in Ref. \cite{Aharon2013}. This method is applicable to a variety of other systems, including hybrid atomic and nanophysics technologies (see Refs.\cite{Bluhm2011,Oelsner2013} and references therein).

Here, we adapt a decoupling scheme, introduced in Ref. \cite{Timoney2011}, such that it is robust against amplitude noise by making use of the multilevel structure of atomic systems and demonstrate that it can be merged with a magnetic sensing protocol to achieve a measurement sensitivity close to the standard quantum limit. Unlike other state-of-the-art magnetometry schemes that either sense signal fields resonant with the pulse rate (pulsed DD, e.g., \cite{Kotler2011}) or the Rabi frequency (continuous DD, e.g., \cite{Hirose2012})  of the decoupling field, the novel magnetometry protocol introduced here relies on the signal to be resonant with a  frequency determined by the decoupling fields' {\em frequencies}.  Today, rf frequencies can easily be stabilized to high precision (e.g., compact commercial atomic clocks provide a relative frequency stability $\Delta\nu/\nu \approx 10^{-12}$) while rf amplitude stability at this level would be challenging to attain. This is in particular true for large Rabi frequencies required for high frequency sensing using continuous or pulsed DD, thus limiting state-of-the-art magnetometry to relatively low frequencies. The sensing scheme introduced here could be tuned to a desired frequency from dc to the GHz range by variation of a static bias magnetic field.

\begin{figure}
\centering {\includegraphics[width=0.7\columnwidth]%, Trim =\trimvalDia]
{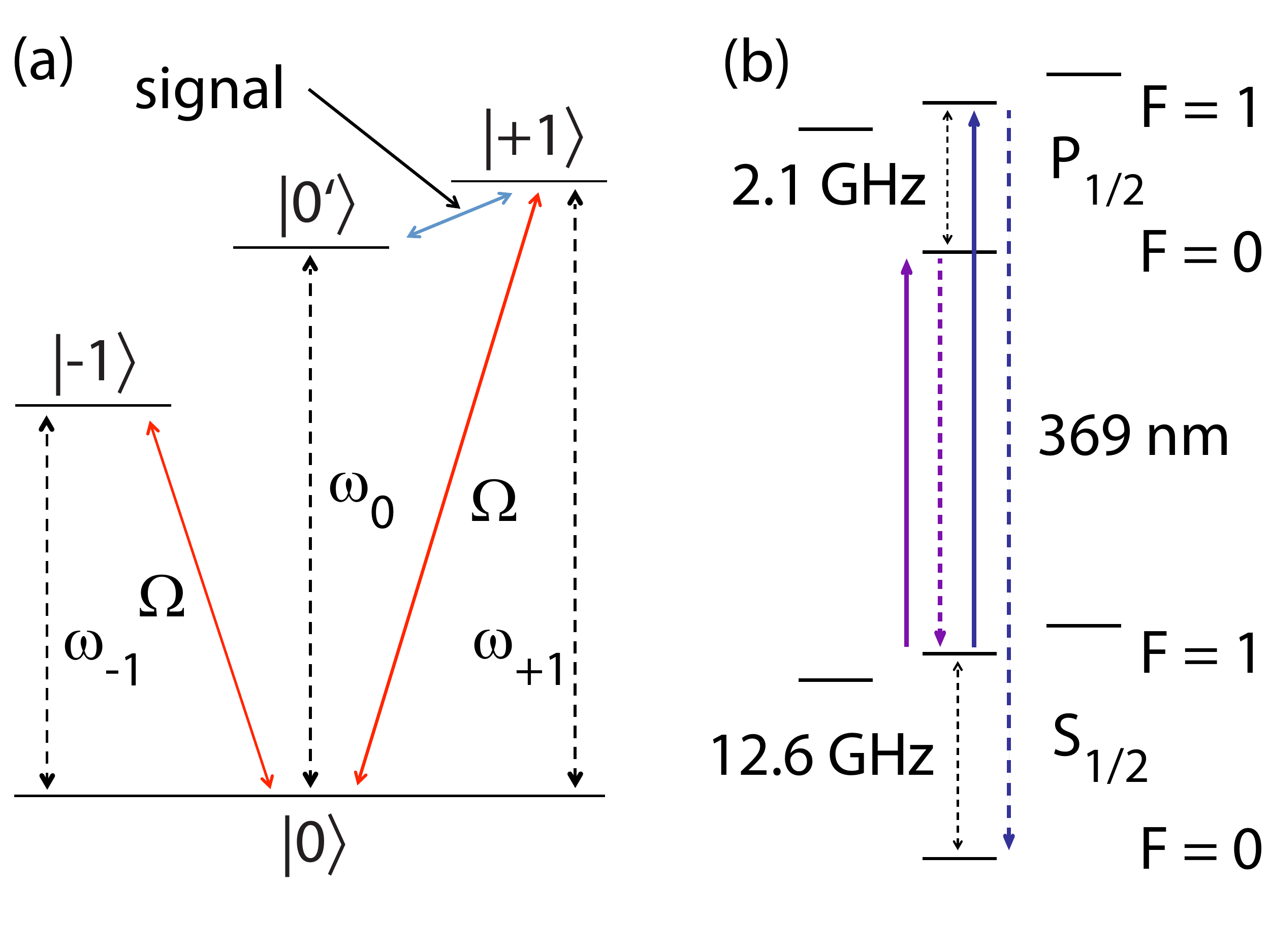}}
\caption{Level scheme for magnetometry. (a) Hyperfine levels for magnetic sensing (not to scale).  The signal field to be measured creates rotations between the states $\ket{0'}$ and $\ket B \equiv \left(\ket {-1} + \ket{+1}\right)/\sqrt 2$, while the microwave fields at frequencies $\omega _{+1}$ and $\omega _{-1}$ and with Rabi frequency $\Omega$ decouple the $\ket B$ state from the orthogonal superposition of the Zeeman sublevels, $\ket D \equiv \left(\ket {-1} - \ket{+1}\right)/\sqrt 2$. The clock transition $\ket 0 \leftrightarrow \ket {0'} $ at frequency $\omega_0$ is insensitive to magnetic field fluctuations to first order. 
(b) Partial $^{171}$Yb$^+$ level structure (not to scale): The four hyperfine states of the electronic ground state are used to implement the level structure shown in  (a). The optical transition near 369 nm is used for Doppler cooling, state preparation, and state-selective readout.}
\label{scheme}
\end{figure}

{\em  Method. ---}  The sensitivity of the magnetometer demonstrated here is dramatically increased by prolonging the coherence time $T_2$ of the quantum states on which the scheme is based. Noise fields acting on the magnetically sensitive bare states $\vert +1 \rangle$ and  $\vert -1 \rangle$ [see Fig.\ref{scheme} (a)]  lead to rapid dephasing of these states. In order to prevent this dephasing, and thus enhance the coherence time $T_2$,  two microwave driving fields are applied as shown in Fig. \ref{scheme} (a). These microwave fields create the dressed state qubit consisting of states $\ket B \equiv  \left( \ket {+1} + \ket {-1} \right)/\sqrt 2$ and $\vert 0' \rangle$. The other two dressed states $\vert u \rangle$ and $\vert d \rangle$ -- superposition states of $\ket D \equiv \left( \ket {+1} - \ket {-1} \right)/\sqrt 2$ and $\ket 0 $ -- are separated from $\ket B$ by an energy gap $ \Omega / \sqrt 2$ (a level scheme of the dressed states is given in the Supplemental Material \cite{supplement}).  Therefore, dephasing of $\ket B$ by ambient fields can occur only, if the ambient field supplies energy at a frequency matching this energy gap. Thus, if the noise field lacks this frequency component, then the coherence time of $\ket B$ is enhanced by orders of magnitude as compared to the bare atomic states \cite{Timoney2011}. State $\ket{0'}$ is insensitive to magnetic fields in first order, and thus, states $\ket B$ and $\ket{0'}$ form a qubit robust against ambient magnetic field fluctuations. However, the $\ket{B} \leftrightarrow \ket{0'}$ transition is sensitive to a particular ac magnetic field -- the signal field to be measured.  

The frequency of the ac signal to be sensed is determined by the frequency difference between the two microwave dressing fields which in turn is set to be of the order of the Zeeman splitting between the atomic states $\ket{0}, \ket {+1}$, and $\ket {-1}$. In order for the ac signal to stimulate a transition between the dressed states  $\ket B$ and $\ket 0'$ it must be resonant (in the bare state picture) with either the transition $\ket{0'} \leftrightarrow \ket {-1}$ or the transition $\ket {0'} \leftrightarrow \ket{+1}$. These two resonances are not degenerate due to the second-order Zeeman shift \cite{Webster2013}.

The level structure shown in Fig. \ref{scheme} (b) is implemented here using hyperfine states of the electronic ground state of a single trapped $^{171}$Yb$^+$ ion. By adjusting the relative phase of the two near-resonant microwave dressing fields, we populate state $\ket B$ \cite{Timoney2011}. Details on the experimental implementation are given in the Supplemental Material \cite{supplement}. Also, Randall {\it et al.}  \cite{Randall2015} give a detailed account of dressed state manipulation using  $^{171}$Yb$^+$ ions.

This  unique opportunity for precise magnetometry  is detailed in the following calculation treating a signal on the $\ket {0'} \leftrightarrow \ket{+1}$ transition. 
The Hamiltonian of the system is 
\bea
H_{sqg}=&& \omega _{0}\ket 0  \bra 0
+\lambda _+(\ket {+1} \bra {+1})
 - \lambda _-(\ket {-1} \bra {-1}) + \nonumber \\
&&\Omega \left( \ket {-1} \bra {0}
e^{i\omega _{-1}t}+e^{i\theta}\ket {1} \bra {0}
e^{i\omega _{+1}t}+H.c. \right)+ \nonumber\\
&&\Omega_g \cos(\lambda _+ t+\phi)\left(\ket {+1} \bra {0'}+H.c. \right),
\eea
where $\omega _0$ is the zero field hyperfine splitting, $\lambda _+$ and $\lambda _-$ are the Zeeman splitting, $\Omega _g$ is the Rabi frequency of the signal field, $\theta$ is the initial phase difference between the two microwave sources at frequencies $\omega _{+1},\omega _{-1}$, and $\phi$ is the initial phase difference between the first microwave source and the signal field (Planck's constant $\hbar$ is set to unity here for convenience).

In the rotating wave approximation and in the interaction picture with respect to the time-independent part after setting $\theta = \pi$ and $\phi = 0,$ we get:
\bea
H =
&&\sqrt{2}\Omega \left( \ket {D} \bra {0}
+H.c. \right)+\sqrt{2}\Omega_g \left( \ket B  \bra {0'}
+H.c. \right)\nonumber \\
=&&\frac{\Omega}{\sqrt{2}}\ket{u}\bra{u}-\frac{\Omega}{\sqrt{2}}\ket{d}\bra{d}+\sqrt{2}\Omega_g
\left( \ket B  \bra {0'} +H.c. \right) ,
\eea 
where $\ket u$ and $\ket d$ are superpositions of states  $\ket 0$ and $\ket D$.
The first two terms shift states $\ket u$ and $\ket d$ (that both contain $\ket D$) away from state $\ket B$, and the second part is the magnetic signal. The interactions created by the two microwave fields with Rabi frequency $\Omega$ decouple the $\ket B$ state from magnetic noise. 

Unlike usual dynamical decoupling, in which the signal field to be measured induces oscillations at the frequency of the pulse sequences or at the Rabi frequency of the continuous field, here oscillations are induced at the frequency difference between a clock transition (that is insensitive to magnetic fields to first order, $\omega _0$ in Fig. \ref{scheme}) and half of the relative detuning of the two microwave frequencies ($\omega _{+1}, \omega _{-1}$). This frequency splitting is close to the Zeeman splitting induced by a dc bias magnetic field. Thus, the frequency of the signal to which the magnetometer is susceptible can be tuned by variation of this bias field to sense a broad range of frequencies. In particular, close to dc  as well as high frequency fields (in the experiments reported here $14.076$ MHz)  can be sensed with a sensitivity close to the standard quantum limit. It should be stressed, that the sensitivity of the  magnetometer is not influenced by possible fluctuations of the static bias magnetic field as these are being suppressed by a factor proportional to  the ratio between the magnetic field fluctuations and the microwave Rabi frequency.

\begin{figure}
\centering {\includegraphics[width=0.8\columnwidth]
	{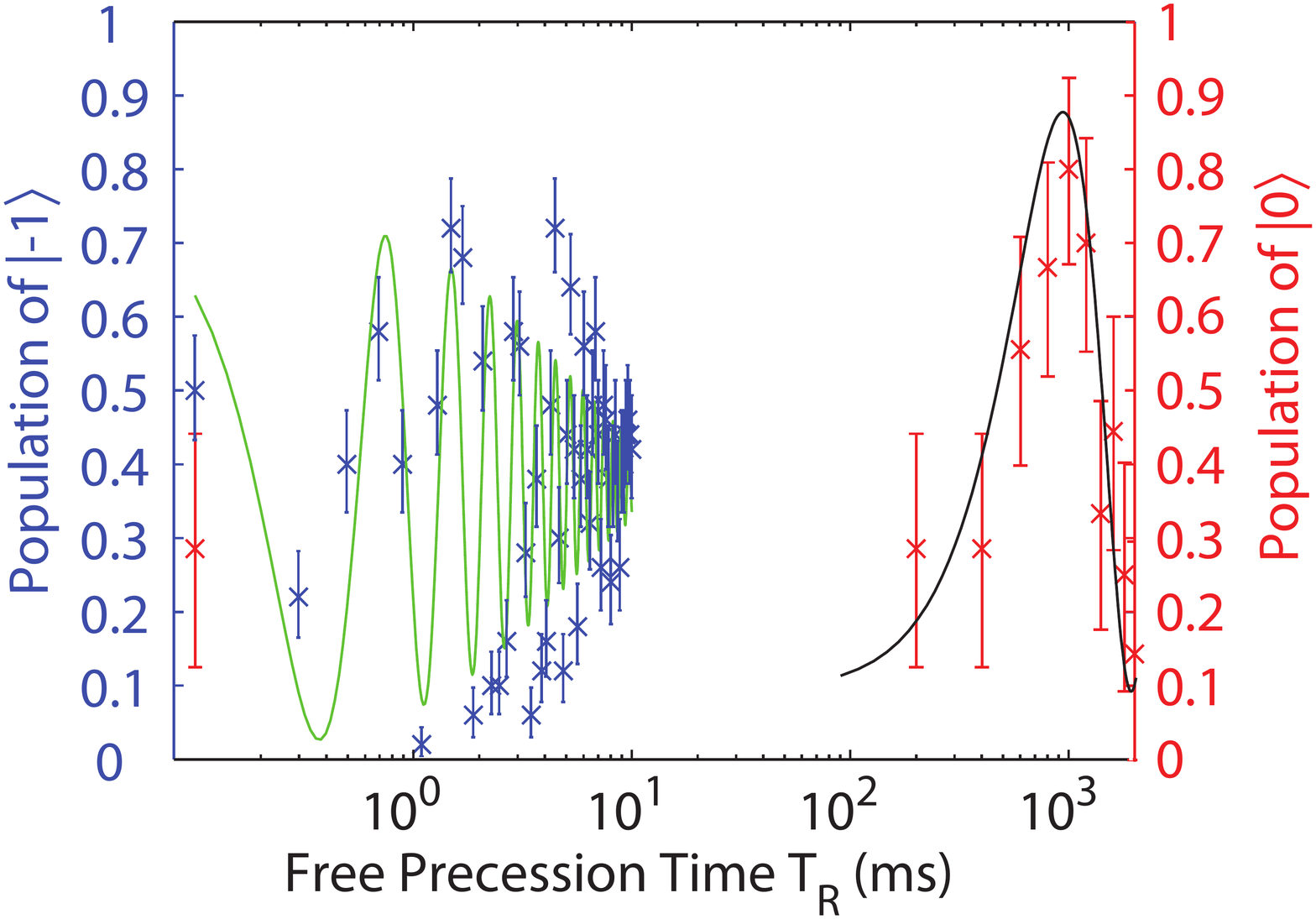}}
\caption{Coherence time of dressed and bare states.   A coherent superposition of $\ket{B}$ and $\ket{0'}$ is prepared and probed after time $T _R$ (red data points and black fit). For comparison, the result of a Ramsey-type experiment between the bare states $\ket{-1} \leftrightarrow \ket{0}$ (blue data points and green fit) is shown. For the dressed states a rf field implements two $\pi/2$-pulses separated by time $T _R$ of free evolution. The rf frequency is slightly detuned from resonance yielding Ramsey oscillations with a period of $1/(0.52 \text{ Hz})$ between $0.1$ ms and $2000$ ms. Each of the $11$ measurement points (dressed states, red data points) consists of $10$ repetitions. In case of the bare states the $51$ measurement points between $0.1$ and $10$ ms are repeated $50$ times each.
The error bars indicate 1 standard deviation.}
\label{fig:ramsey_oszib0'}
\end{figure}

{\em Ramsey oscillations and magnetometry sensitivity. ---}  The coherence time of the bare atomic states and of the dressed state qubit ($\ket{B}$ and $\ket{0'}$), respectively, is tested by a Ramsey-type experiment. After creating a superposition of the two magnetically sensitive $m _F=\pm 1$ bare states, their coherence is rapidly lost  due to fluctuating ambient magnetic fields. The bare states are characterized in this experiment by a coherence time of 5.3 ms whereas the experimental result in Fig. \ref{fig:ramsey_oszib0'} shows that coherence can be preserved for more than $2000$ ms when using the dressed state qubit. Thus, the coherence is preserved for a time almost 3 orders of magnitude longer than the dephasing time of the atomic states $\ket{-1}$ and $\ket{+1}$ (see supplemental material for a discussion on the limits of this method \cite{supplement}).

\begin{figure}
	\centering {\includegraphics[width=0.8\columnwidth]%, Trim =\trimvalDia]
		{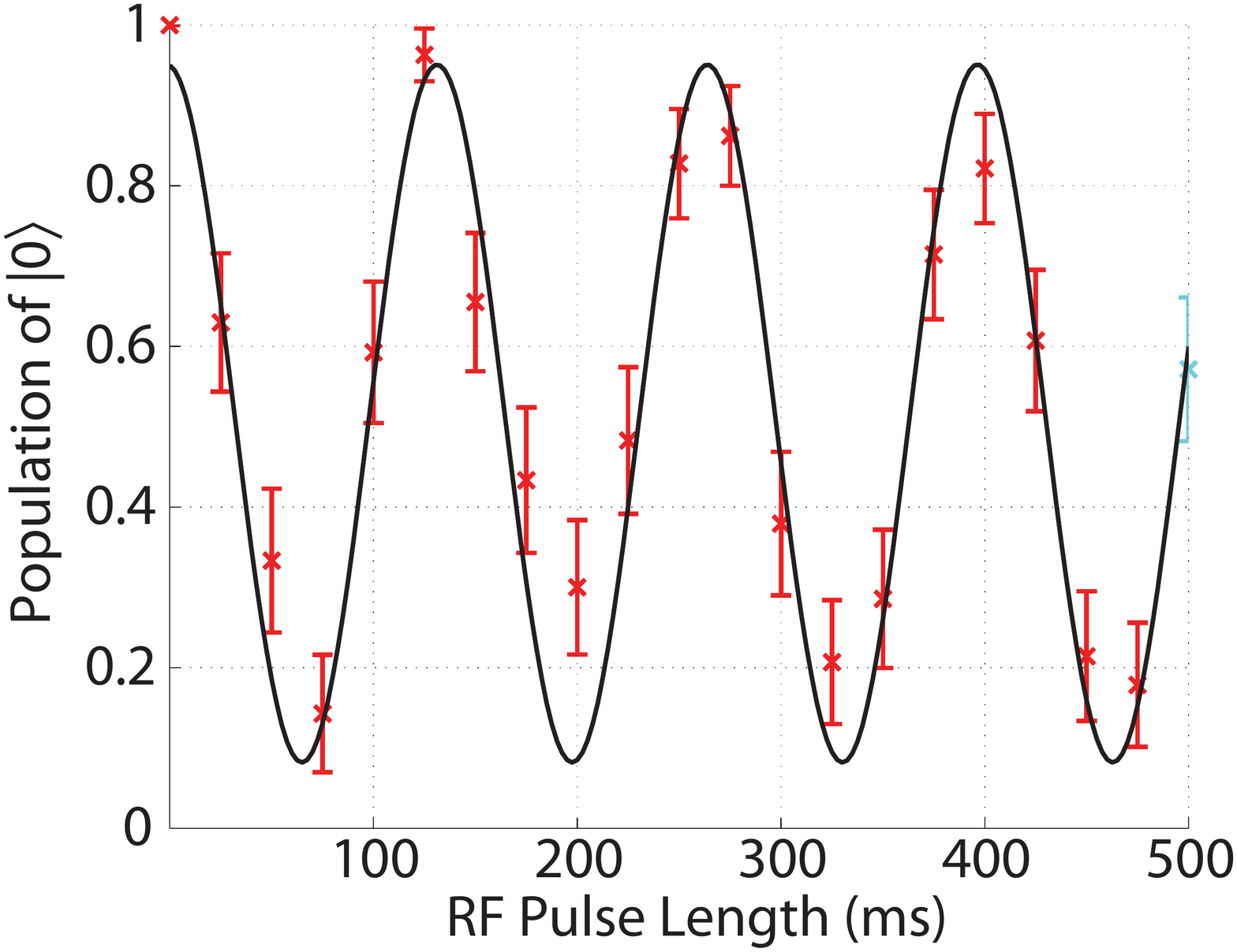}}
	\caption{Qubit rotation between coherence protected states $\ket{B}$ and $\ket{0'}$. Rabi oscillations take place between $0.1$ ms and $500$ ms with a Rabi frequency $\Omega_g = 2\pi \times (7.54\pm0.12)$ Hz. Each of the $21$ measurement points consists of $30$ repetitions. The solid line is a fit to the data yielding $\Omega_g$. 
%The experimental data point for $T = 500$ ms is presented in cyan color coding. 
The error bars indicate 1 standard deviation.}
	\label{fig:rabi_oszib0'_1s}
\end{figure}

In order to measure the sensitivity of the magnetometer, we apply an rf field set to resonance with the  $\ket{0'}\leftrightarrow\ket{+1}$ transition  to induce Rabi oscillations (Fig. \ref{fig:rabi_oszib0'_1s}) with  frequency  $\Omega _g$ between the dressed state $\ket{B}$ and state $\ket{0'}$ \cite{Webster2013}. The population $P(T)$ of state $\ket{-1}$  (which corresponds to the population of $\ket{B}$) after application of the rf pulse is mapped onto state $\ket{0}$ and the population of state $\ket{0}$ is optically readout \cite{supplement}. Rabi oscillations driven by rf radiation are sustained for $500$ ms in this particular example, demonstrating the extended coherence of magnetically sensitive states after dressing them with microwave fields. Rabi oscillations for more than 2 sec were also recorded (not shown).

\begin{figure}
\includegraphics[width=1.0\columnwidth]{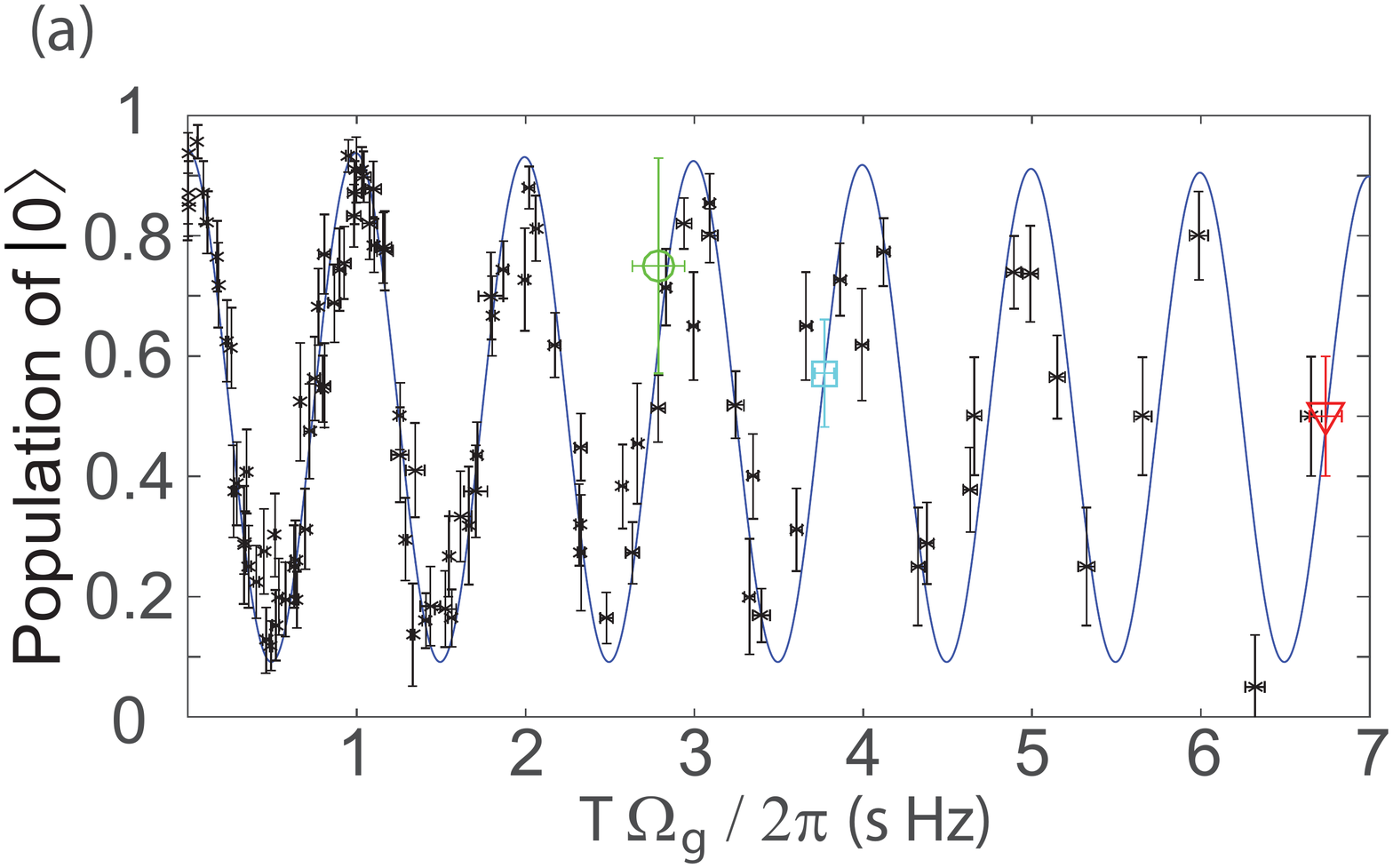}
\includegraphics[width=1.0\columnwidth]{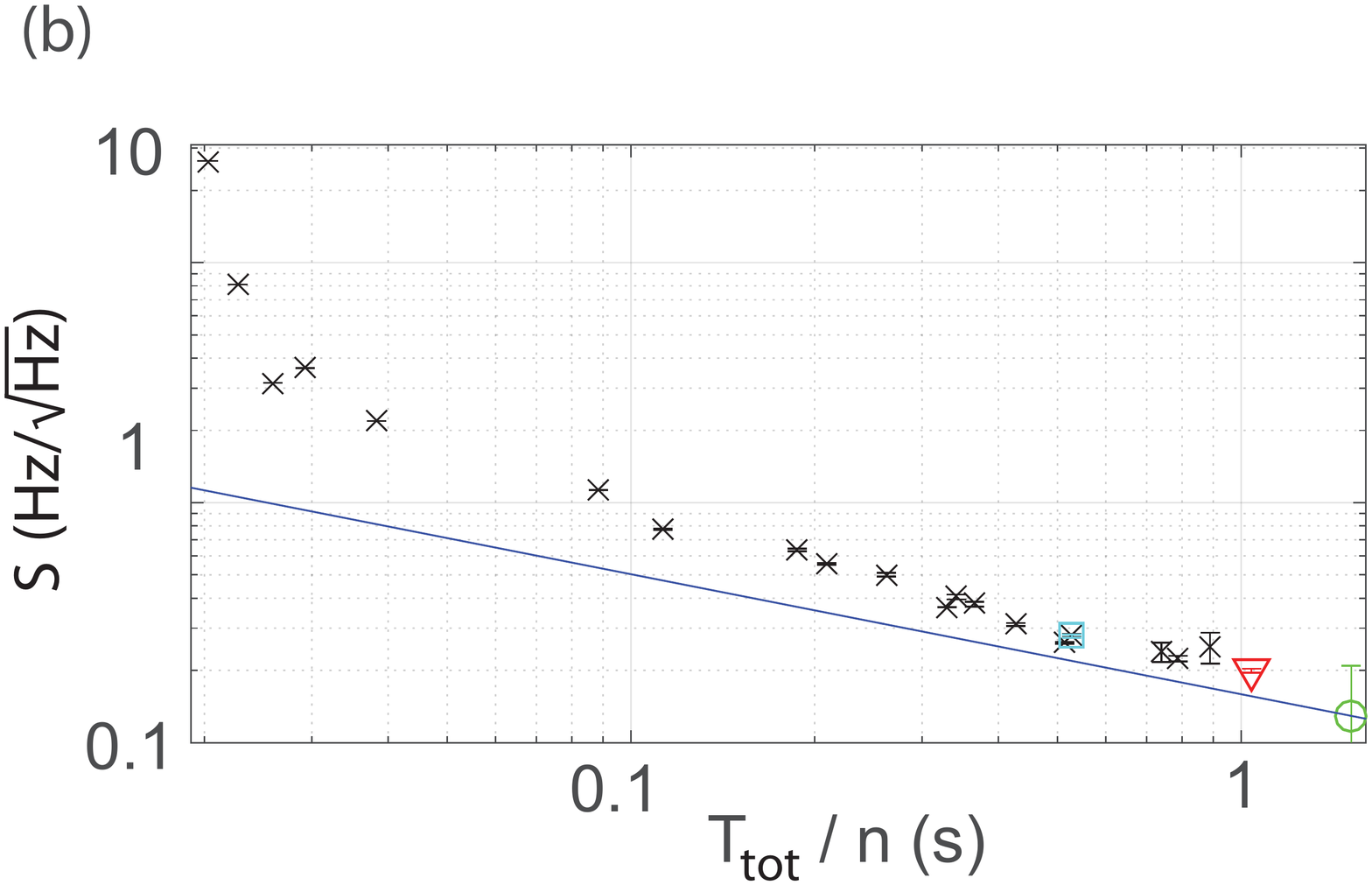}
\caption{Sensitivity of the dressed state magnetometer. (a) Rabi signal measured for various pulse durations $T$ and different values of $\Omega _g$ plotted versus $\varphi = \Omega _g \times T$.  The range of $\Omega_g$ extends from  $2\pi\times 2$ Hz to $2\pi\times 3336$ Hz \cite{supplement}. The derivative of the blue fitted curve gives the slope $\partial P(\varphi)/\partial \varphi$. Each data point is the average of $n$ repetitions. (b) Sensitivity $S$ for variation in $\Omega_g$ as a function of the total measurement time $T_{tot}/n = (T+T_{add})$ where  $T_{add}$ is the additional time needed for state preparation, read out, and laser cooling.  
The sensitivity derived for three exemplary data points in a) is shown with different color coding and marker symbol: For an experimental time $T = 500$ ms and $\Omega_g=2 \pi\times (7.54\pm0.12)$ Hz (see Fig. 3), a sensitivity of $S = (0.278 \pm 0.017)$ $ \mbox{Hz}/\sqrt{\mbox{Hz}}$ is achieved (cyan). For  $T = 1000$ ms and $\Omega_g=2 \pi\times(6.74\pm0.10)$ Hz a sensitivity of $S = (0.200 \pm 0.004)$ $\mbox{Hz}/\sqrt{\mbox{Hz}}$ is derived (red triangle). The best sensitivity $S = (0.130 \pm 0.036)$ $\mbox{Hz}/\sqrt{\mbox{Hz}}$ is reached for an experimental time $T = 1500$ ms and $\Omega_g=2 \pi\times (1.86\pm0.10)$ Hz (green circle).  The solid blue line indicates the standard quantum limit given by $1/\sqrt{T}$. 
}
\label{fig:Sens_sum}
\end{figure}

The sensitivity for a variation of $\Omega_g$ is given by 
\begin{equation}
\delta \Omega_g =\frac{\Delta P}{\left| \frac{\partial P(\varphi)}{\partial \varphi}\right|  T}
\label{eq:deltaOmega}
\end{equation} 
where $\Delta P$ is the standard deviation  associated with the population measurement  after time $T$ (error bar in Fig.\ref{fig:Sens_sum} (a)) for phase $\varphi =\Omega_g T$, which is the product of Rabi frequency $\Omega_g$ and the duration $T$ of an rf pulse during which Rabi oscillations are observed (see supplemental material for more details \cite{supplement}). In Fig.\ref{fig:Sens_sum}(a), we show the population $P(\varphi)$ of state $\ket{0}$  for different durations $T$ of rf pulses and various values of $\Omega_g$ (the exact values of $\Omega_g$ are given in the supplemental material \cite{supplement}). The factor 
$ \frac{\partial P(\varphi)}{\partial \varphi}$ in Eq. \ref{eq:deltaOmega} is the slope of the blue fitted function in Fig. \ref{fig:Sens_sum} (a).

The shot-noise-limited sensitivity $S$ for the measurement of $\Omega_g$ is given by \cite{Itano93,Taylor2008}
\begin{equation}
S=\delta\Omega_g\sqrt{T_{tot}} .
\label{eq:S}
\end{equation} 
The quantity $S$ (in units of $\mbox{Hz}/\sqrt{\mbox{Hz}}$) characterizes the minimal change in $\Omega_g$ that can be discriminated within a total experimental time of one second.
Here,  $T_{tot} = n ( T + T _{add})$ is the total time needed  for $n$ repetitions of the measurement with Rabi oscillation time $T$.  $T _{add}$  is the additional time needed -- for state preparation,  readout,  and for cooling of the ion -- when recording the result of a single repetition of a given data point. During time $T _{add}$ the magnetometer is not sensitive to the signal. 
For short $T_{tot}$, the sensitivity is limited by the constant overhead in time, $T_{add}$. 
When $T$ becomes longer, the relative contribution of $T _{add}$ to the total measurement time $T_{tot}$ decreases and the experimental sensitivity approaches the standard quantum limit given by $S_Q=  1/\sqrt{T}$  (blue line in Fig.\ref{fig:Sens_sum} (b); see also supplemental material \cite{supplement}). 

Exemplary, the sensitivity of three data points is emphasized in Fig.\ref{fig:Sens_sum} (b) with different color coding and marker symbols that correspond to the data points of the same color in Fig.\ref{fig:Sens_sum} (a). 
The green circle in Fig. \ref{fig:Sens_sum} (b) corresponds to a sensitivity of $4.6$ pT $/\sqrt{\mbox{Hz}}$ at a signal frequency near 14 MHz. 

{\em Summary and outlook.---}
A novel method for magnetometry is presented that is robust against amplitude fluctuations of the decoupling fields. We have demonstrated this method using a single atomic ion confined to a spatial region of order (20 nm)$^3$ and showed a record sensitivity for magnetic fields near 14 MHz. This method can be applied to operate in a broad range of frequencies from dc to GHz. Thus, using suitable ion traps, such a single atom sensor can be brought into the vicinity of  samples to be investigated and can be used to image with unprecedented sensitivity and nanometer spatial resolution magnetic fields over a broad range of frequencies. A microfabricated so-called stylus trap would be particularly well suited to bring a single atom close to a sample to be investigated \cite{Arrington2013}.  Furthermore, this method can be adapted to be used in other atomic systems and in ensembles increasing the signal-to-noise-ratio of magnetometry in a broad range of platforms. 
This technique  reaches a high magnetic sensitivity even in the presence of an offset field. Using two or more ions, this method can also be used to detect magnetic field {\em gradients} with high resolution in space and amplitude. 

This work was supported by the Career Integration Grant (CIG) IonQuanSense, the Israeli Science foundationan, by the Bundesministerium f\"ur Bildung und Forschung (FK 01BQ1012), an Alexander von Humboldt Professorship, the EU Integrating Projects DIADEMS and SIQS, the EU STREP projects EQUAM and iQIT,  and the ERC Synergy grant BioQ.  

*Corresponding author:\\ christof.wunderlich@uni-siegen.de

\newpage

\section*{Supplemental Material}
{\bf Experimental Procedures --}
The experiments are carried out using a  single Doppler cooled $^{171}$Yb$^{+}$ ion confined in a 2-mm-sized Paul trap. The ion is localized in a spatial region of about (20 nm)$^3$. The electronic ground state S$_{1/2}$ splits up into hyperfine states with total angular momentum  quantum number F = 0 (state $\ket{0}$) and F = 1 ( states $\ket{-1}$ ,  $\ket{0'}$ and  $\ket{+1}$), where  the magnetic quantum number m$_F$ is used as a label (main text, Fig.1). The degeneracy of the Zeeman levels is lifted by a static bias magnetic field. The second order Zeeman effect shifts states  $\ket{-1}$ and $\ket{+1}$ by different absolute values. The ion is initialized in state $\ket 0$ by optical pumping on the S$ _{1/2}$, F = 1 $\leftrightarrow$ P$ _{1/2}$, F = 1 resonance (main text, Fig.1b). State selective detection at the end of a measurement sequence is achieved by applying laser light to drive the S$ _{1/2}$, F = 1 $\leftrightarrow$ P$ _{1/2}$, F = 0 resonance and detecting scattered resonance fluorescence. After initializing the ion, the preparation  of a desired dressed state is done via an incomplete stimulated Raman adiabatic passage (STIRAP) sequence \cite{Bergmann1998,Vitanov2001} with microwave fields. The effectiveness of the STIRAP process for creating dressed states with microwave fields is investigated in Ref. \cite{Timoney2011}.

\begin{figure*}[htpb]
	\centering {\includegraphics[Trim={20 180 60 180},width=1.65\columnwidth]{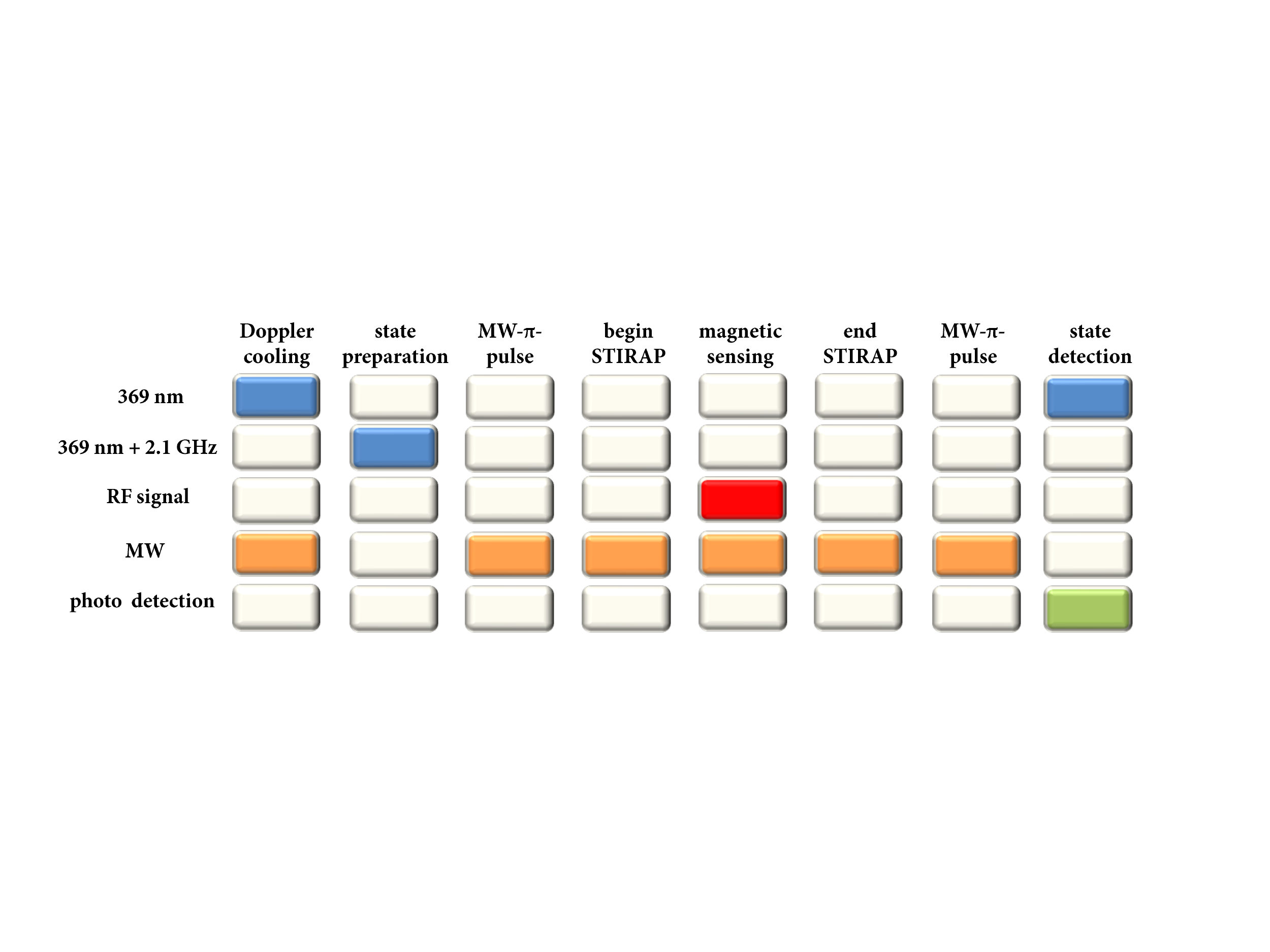}}
	\caption{Experimental sequence to take one datum using a single trapped $^{171}$Yb$^+$ ion. The labels of each row indicate the elements that are turned on and off in order to achieve the action indicated above each column. Laser light near 369 nm is used for Doppler cooling (first step) and state selective detection (last step). Laser light near 369 nm but detuned by 2.1 GHz relative to the light field above serves for initial preparation in state $\ket{0}$.  Resonant microwave fields serve for transferring population between states $\ket{0}$ and $\ket{0'}$ (steps 3 and 7). Two microwave fields are used to prepare dressed states (step 4) and to transfer them back to the the bare state basis (step 6) using stimulated raman adiabatic passage (STIRAP). The RF signal introduces a transition between dressed states. Resonance fluorescence is detected using a photomultiplier. }
	\label{scheme_supp}
\end{figure*}

For creating these dressed states, two microwave fields are used with Gaussian amplitude envelopes shifted in time relative to each other driving the $\ket{-1} \leftrightarrow \ket{0}$ and the $\ket{+1} \leftrightarrow \ket{0}$ transitions, respectively. Microwave fields are generated using two phase locked RF sources adjustable in frequency, amplitude, and phase. Depending on the initial state, on the relative phase between the two microwave fields, and on the order of the pulse sequence, the incomplete STIRAP sequence prepares different dressed states, and ends after completion of the sequence in different final atomic states. In order to prepare the dressed state $\ket{B} $, a microwave $\pi$-pulse transfers the population from the initial state $\ket 0$ to the atomic state $\ket{+1}$. The first pulse of the STIRAP sequence drives the $\ket{-1} \leftrightarrow \ket{0}$ resonance and the second microwave field the $\ket{+ 1} \leftrightarrow \ket{0}$ resonance. A relative phase of $\pi$ between the two microwave dressing fields leads to a transfer of atomic population to the dressed state $\ket{B}$. In these experiments,  the maximum Rabi frequency of the two microwave fields, $18 \times 2\pi$ kHz$\leq \Omega \leq 40 \times 2\pi$ kHz. To keep the dressed state $\ket B$ populated, amplitudes and phases of the two microwave dressing fields are kept constant. When the STIRAP sequence is completed, the population is brought back to the atomic basis to state $\ket{-1}$. Before probing the final state, a microwave $\pi$-pulse swaps the population of $\ket{-1}$ and $\ket{0}$. Now it is possible to distinguish the population of states $\ket B$ and $\ket{0'}$ when detecting scattered resonance fluorescence on the S$ _{1/2}$, F = 1 $\leftrightarrow$ P$ _{1/2}$, F = 0 resonance (main text, Fig.1b). The experimental sequence outlined above is summarized in Fig. 1 of this supplement.

For inducing Rabi oscillations between the dressed state $\ket B$ and the atomic state $\ket{0'}$ there exist different ways: Using two RF fields on resonance with the transitions $\ket{0'} \leftrightarrow \ket{\pm1}$, or as shown in \cite{Webster2013}, using one RF field on the resonance of $\ket{0'} \leftrightarrow \ket{+1}$ or $\ket{0'} \leftrightarrow \ket{-1}$ is sufficient to initiate Rabi oscillations between  $\ket B$ and $\ket{0'}$, as long as $\Omega_g \ll \Omega$ and $\Omega_g$ is much smaller then the detuning of the two microwave dressing fields. %This reduces the effective Rabi frequency by factor two ($\Omega_g/\sqrt{2}$), compared to the method used in \cite{Timoney2011} \intern{($\sqrt{2}\Omega$) - ja}. 
For the experimental data shown here the frequency of the RF field is applied on the resonance of the transition $\ket{0'} \leftrightarrow \ket{+1}$ (main text, Fig.1a). The RF pulse is applied during the hold in evolution of the STIRAP pulse sequence (the microwave dressing fields are held at constant amplitude) once the dressed state $\ket{B}$ is populated. After the RF Rabi pulse, the STIRAP sequence is completed and the population of state $\ket{0}$ is probed. When the RF Rabi pulse populates state $\ket{0'}$, the second part of the STIRAP sequence has no effect and population will remain in $\ket{0'}$ resulting in a bright signal during detection. Rabi oscillations are shown in Fig. 3 of the main text. Taking into account a possible decay of the contrast due to dephasing, the signal takes the form of $P(T)=\frac{1}{2}[1+ f(T) cos(\Omega _g T)]$. For typical data as shown in Fig. 3 and Fig. 4 of the main text, $f(T)\approx$ constant. 

For the Ramsey measurement shown in Fig. 2 (main text) the STIRAP pulses are characterized by these parameters: pulse separation of $5/f _{\Omega} = 135$ $\mu$s, pulse width of $8.35/f _{\Omega} = 224$ $\mu s$ and discrete time-increments $\Delta t = \frac{1}{10f _{\Omega}} = 2.68$ $\mu$s, with $\Omega = 2\pi\times37.27$ kHz. 
%Each of the $11$ measurement points consists of $10$ repetitions. 
The microwave frequency on the $\ket{+1}\leftrightarrow  \ket{0}$ resonance is set to $12.6531$ GHz and on the $\ket{-1} \leftrightarrow  \ket{0}$ resonance it is $12.6325$ GHz. A static magnetic field of $B = 0.732 $ mT defines a quantization axis. In case of the Ramsey measurement with the bare states, two $\pi/2$-pulses on the $\ket{-1} \leftrightarrow \ket{0}$ resonance are applied. The microwave field is slightly detuned from the resonance at $12.6328$ GHz. 
%$51$ measurement points between $0.1$ ms and $10$ ms are repeated $50$ times.

The experimental parameters for the data shown in Figs. 3 and 4 are as follows:
The parameters of the STIRAP pulses are defined by $f _{\Omega}=\Omega/(2\pi)$: pulse separation of $5/f_{\Omega} = 278$ $\mu s$, pulse width of $8.35/f_{\Omega} = 464$ $\mu s$ and discrete time-increments $\Delta t = \frac{1}{10f_{\Omega}} = 5.56$ $\mu s$, with $\Omega = 2\pi\times18$ kHz. Each of the $21$ measurement points in Fig. 3 consists of $n=30$ repetitions. The microwave frequency on the $\ket{+1} \leftrightarrow \ket{0}$ resonance was $12.6569$ GHz and on the $\ket{-1} \leftrightarrow \ket{0}$ resonance was $12.6287$ GHz. A static magnetic field of $B = 1$ mT defines a quantization axis. The data in Fig. 4 is obtained by measuring Rabi oscillations for various pulse durations $T$, Rabi frequencies $\Omega_g$, and number of repetitions $n$ (see table \ref{table1}).

In order to find an expression for the standard quantum limit of the sensitivity, $S_Q$, we start from the expression for the minimal detectable change $\delta \Omega_g$ given in equation 3 of the main text 
that we repeat here for convenience:  
\begin{equation}
\delta \Omega_g =\frac{\Delta P}{\left|\frac{ \partial P(\varphi)}{\partial \varphi}\right|  T} \ .
\label{eq:deltaOmega_supp}
\end{equation} 
For evaluating the experimental sensitivity shown in Fig. 4 of the main text we used the experimental value of $\Delta P$ for a given data point. The ideal {\em theoretical} uncertainty in determining $\Delta P$ is given by 
\begin{equation}
\Delta P=\frac{\sqrt{\langle P \rangle (1-\langle P \rangle)}}{\sqrt{n}}
\label{eq:DeltaP}
\end{equation} 
with $P = \ket{0}\bra{0}$ \cite{Itano93}. Using $\langle P \rangle=\cos^2(\varphi/2)$ with $\varphi=\Omega_g T$, equation \ref{eq:deltaOmega_supp} gives the expression   
\begin{equation}
\delta \Omega_g =  \frac{1}{T} \sqrt{\frac{T_s}{T_{tot}}} \ .
\label{eq:deltaOmega2}
\end{equation} 
Here, in addition, we made use of the relation $T_{tot}=n T_s=n(T+T_{add})$ to replace $1/\sqrt{n}$ by 
$\sqrt{T_s/T_{tot}}$.
For an experiment with the total experimental time $T_{tot}$ (averaging over $n$ measurements to obtain a reading of the magnetometer), the time $T_s $ needed for a {\em single} measurement is the sum of the Rabi oscillation time $T$ and the additional time $T_{add}$ for cooling of the ion, preparation of its initial state, and state readout.
If $T_{add}$ is small compared with $T$ (i.e., for $T_{s}\rightarrow T $), then the theoretical limit for the sensitivity is obtained as 
\begin {equation}
S_Q=\delta \Omega_g \sqrt{T_{tot}}  =  \frac{1}{\sqrt{T}} \ .
\end{equation}

 \renewcommand{\arraystretch}{1.5}
 \begin{table}[htpb]
 	\centering 
 	\begin{tabular}{cp{0.02 cm}cp{0.02 cm}cp{0.3 cm}cp{0.25 cm}c} %{cp{0.5 cm}ccp{0.5 cm}c}
 		%\toprule
 %		\textbf{$\Omega_g$ ($2\pi\times$ Hz)} &&\textbf{$\Delta\Omega_g$ ($2\pi\times$ Hz)} &&\textbf{n}&& \textbf{T (s)}&&\textbf{T$_{add}$ (s)}\\
  		$\Omega_g$ ($2\pi\times$ Hz) &&$\Delta\Omega_g$ ($2\pi\times$ Hz) &&n&&T (s)&&T$_{add}$ (s)\\
 		\hline
 		\hline
 	\end{tabular}
 	\vspace{2mm} 
 	\renewcommand{\arraystretch}{0.5}
 	\begin{tabular}[b]{lp{1.0 cm}lp{0.8 cm}lp{0.15 cm}lp{0.15 cm}lp{0.15 cm}}
 		3336.37&&24.63&&20&&0.0008&&0.019\\
 		431.58&&13.24&&20&&0.0043&&0.018\\
 		414.23&&6.55&&20&&0.0090&&0.017\\
 		230.18&&5.90&&20&&0.0095&&0.020\\
 		218.83&&3.21&&20&&0.0170&&0.021\\
 		34.84&&3.38&&10&&0.0500&&0.039\\
 		20.17&&1.95&&10&&0.0900&&0.022\\
 		18.79&&0.51&&10&&0.1667&&0.020\\
 		13.62&&0.56&&10&&0.1700&&0.040\\
 		7.70&&0.27&&10&&0.2400&&0.022\\
 		12.52&&0.17&&40&&0.3000&&0.028\\
 		18.21&&0.19&&20&&0.3200&&0.020\\
 		18.21&&0.19&&20&&0.3467&&0.020\\
 		5.53&&0.16&&10&&0.4000&&0.028\\
 		2.65&&0.16&&10&&0.4750&&0.038\\
 		5.53&&0.16&&10&&0.5000&&0.028\\
 		7.54&&0.12&&30&&0.5000&&0.028\\
 		3.14&&0.14&&10&&0.7000&&0.040\\
 		5.67&&0.18&&5&&0.7500&&0.039\\
 		6.65&&0.06&&25&&0.8500&&0.039\\
 		6.74&&0.10&&25&&1.0000&&0.039\\
 		1.86&&0.10&&5&&1.5000&&0.015\\
 	\end{tabular}
 	\caption{Experimental parameters of the measurements shown in Fig. 4 of the main text. 1. column: Rabi frequency $\Omega_g$, 2. column: 
standard deviation $\Delta\Omega_g$, 3. column: number of repetitions $n$, 4. column: Rabi oscillation time $T$, 5. column: Time $T_{add}$ needed for state preparation, cooling and read-out. The order of the values listed here (top to bottom) corresponds to Fig. 4 (b) with ascending phase $\varphi =\Omega_g T$. }
 	\label{table1}
 \end{table}

\vspace{1cm}
\begin{figure}[htpb]
	\centering {\includegraphics[width=0.7\columnwidth]%, Trim =\trimvalDia]
		{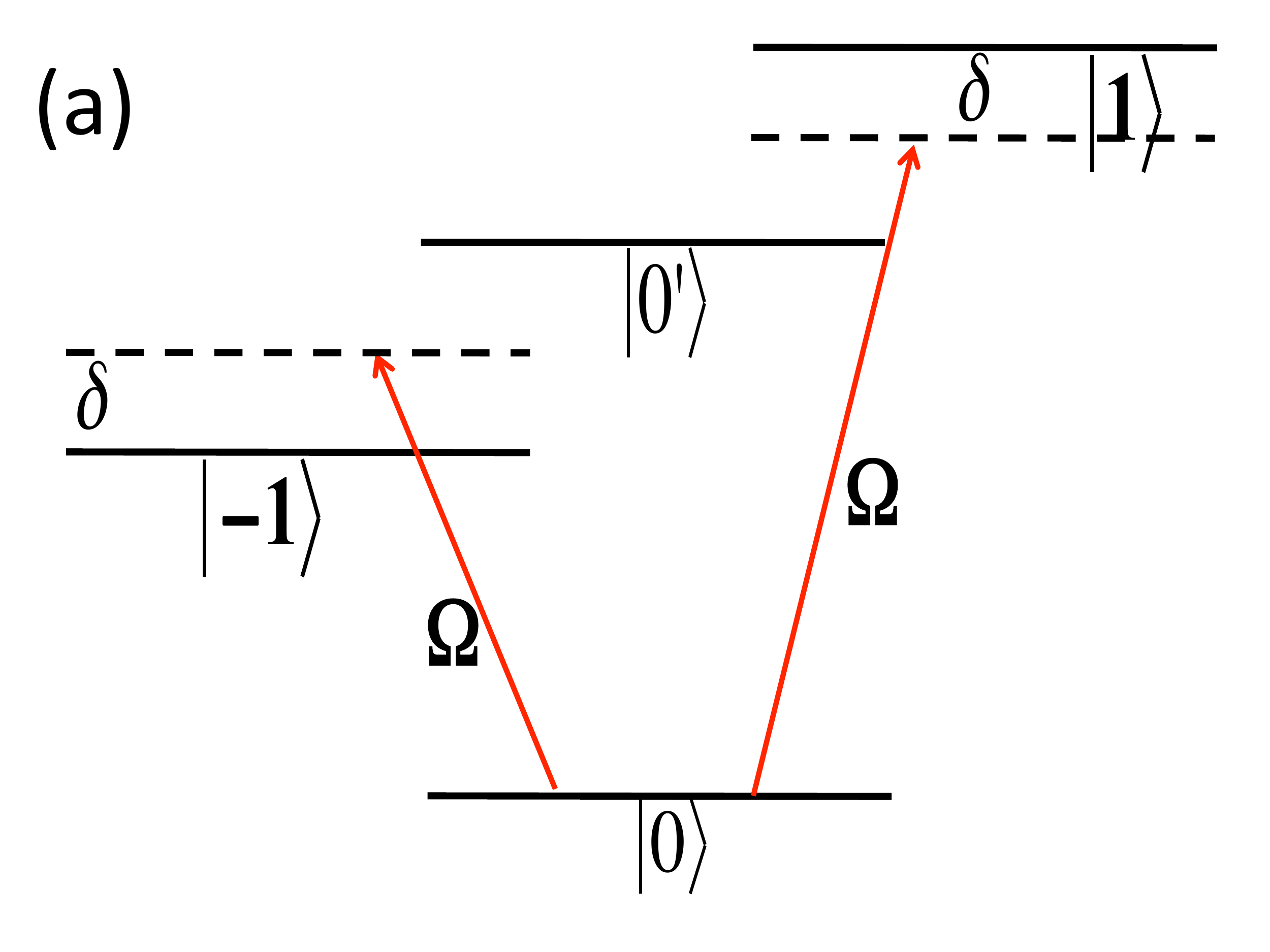} 
		\includegraphics[width=0.7\columnwidth]%, Trim =\trimvalDia]
		{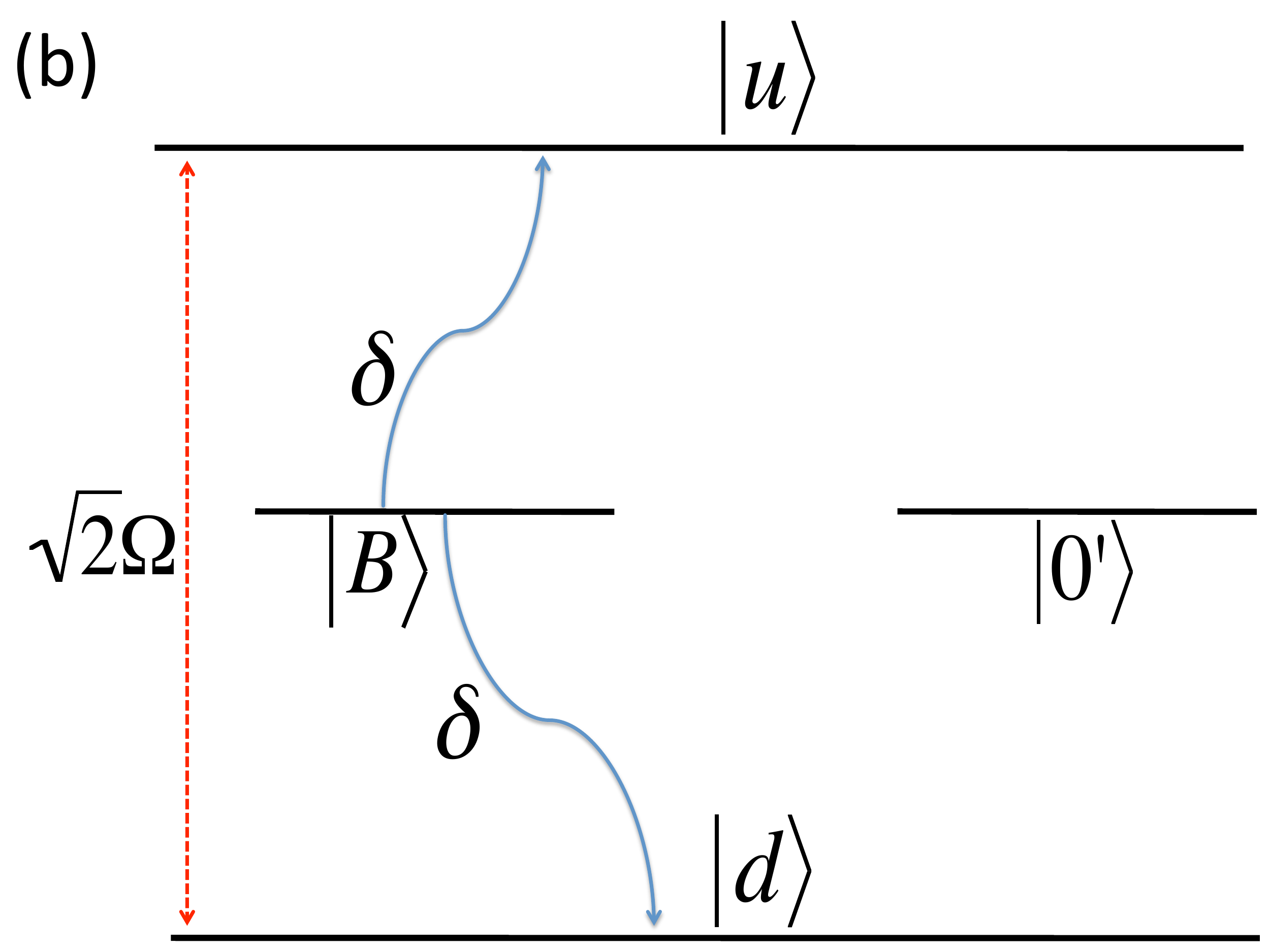}
		%\includegraphics[width=0.45\columnwidth]%, Trim =\trimvalDia]
		%{supp_theory_c} \includegraphics[width=0.45\columnwidth]%, Trim =\trimvalDia]
		%{supp_theory_d}
		}
	\caption{  (a) Magnetic field induced detuning $\delta(t)$ in the bare state basis.  (b) Frequency components of the time dependent field (causing a level shift $\delta(t)$ in the bare basis) that match the gap between the dressed states couple state $\ket{B}$ to states  $\vert u \rangle$ and $\vert d \rangle$.  Due to symmetry, the  $T_2$ time of the sensing qubit $\{\ket{B},\ket{0'} \}$ is not affected by this coupling. }
	\label{scheme_theory}
\end{figure}

{\bf Theoretical Considerations --}
The  magnetic field to which the sensor is exposed can be decomposed into a mean value (the bias field) and fluctuations around this mean value characterized typically by a broad power spectral density and with zero time average. 
The magnetic field acting on magnetically sensitive states $  \ket {+1}$ and $ \ket {-1}$  leads to a  shift of magnitude $\delta$  of these states $  \ket {+1}$ and $ \ket {-1}$ (Fig. \ref{scheme_theory} a)). Thus, if the field fluctuates, it will change the relative phase between these bare states when a superposition state is created. A broad range of 
frequency components contribute to this dephasing.  

However, when the bare states are dressed by microwave fields, as described in the main text, the energy shift $\delta$ is transformed into a coupling (proportional to $\delta$) between the dressed states $\vert B \rangle$ and states  $\vert u \rangle$ and $\vert d \rangle$ (Fig. \ref{scheme_theory} b)). The dressed states $\vert u \rangle$ and $\vert d \rangle$ are well separated in energy from the  $\vert B \rangle$ state. Now, state $\vert B \rangle$ is protected from dephasing, since only specific magnetic field components  at or around the frequency corresponding to the separation $\Omega/\sqrt{2}$ between the dressed states contribute to this coupling, and consequently the transitions  $\vert B \rangle \leftrightarrow \vert u \rangle$ and $\vert B \rangle \leftrightarrow \vert d \rangle$ induced by magnetic noise are  suppressed (here, all AC magnetic components different from the one to be measured are referred to as noise). Therefore, dephasing is strongly suppressed as  compared to the case of bare states. (The Rabi frequency of the microwave dressing fields is indicated by $\Omega$.) 

We use the qubit composed of states $\ket{B}$ and $\ket{0'}$ for magnetometry and now consider the limits for the coherence time of this qubit. The coupling of $\ket{B}$ to states $\vert u \rangle$ and $\vert d \rangle$, respectively, causes level shifts of state $\ket{B}$ and thus could dephase the qubit. However, these level shifts in the dressed state basis  are calculated as  $\frac{\delta^2}{\Omega/\sqrt{2}}$ and $-\frac{\delta^2}{\Omega/\sqrt{2}}$, respectively. That is, they are of equal magnitude but opposite sign and therefore cancel each other in all orders of the magnetic field. 

However, the qubit is composed of the two states $\ket B$ and $ \vert 0' \rangle$, and it is in fact state  $ \vert 0' \rangle$ that is the  limiting factor for the qubit's coherence time. This state is affected by the second order Zeeman effect, and fluctuations in the energy of state  $\vert 0' \rangle $ can dephase the qubit and therefore limit the $T_2$ time of the sensing qubit, and in turn the sensitivity of the magnetometer.

So far, we have considered level shifts that could cause dephasing. As explained above, the energy of state $\ket B$ does not change as a consequence of magnetic noise fields and there is no dephasing of this qubit state. However, due to the coupling of state $\ket{B}$ to states $\ket{u}$ and $\ket{d}$,  state $\ket{B}$ is ``contaminated««  by $\ket u$ and $\ket d$ with an amplitude   $\frac{\delta}{\Omega/\sqrt{2}}.$ Meaning, the eigenstate will change due to the effect of the noise to $\ket B_{new }  \approx \ket B + \frac{\Omega}{\sqrt{2} \delta} \ket u -\frac{\Omega}{\sqrt{2} \delta} \ket d$. Due to this contamination the coupling to the signal will be reduced by the same amount and will decrease the sensitivity. 

\vspace{1cm}
\begin{figure}[!tb]
	\centering {
		 \includegraphics[width=0.85\columnwidth]%, Trim =\trimvalDia]
		{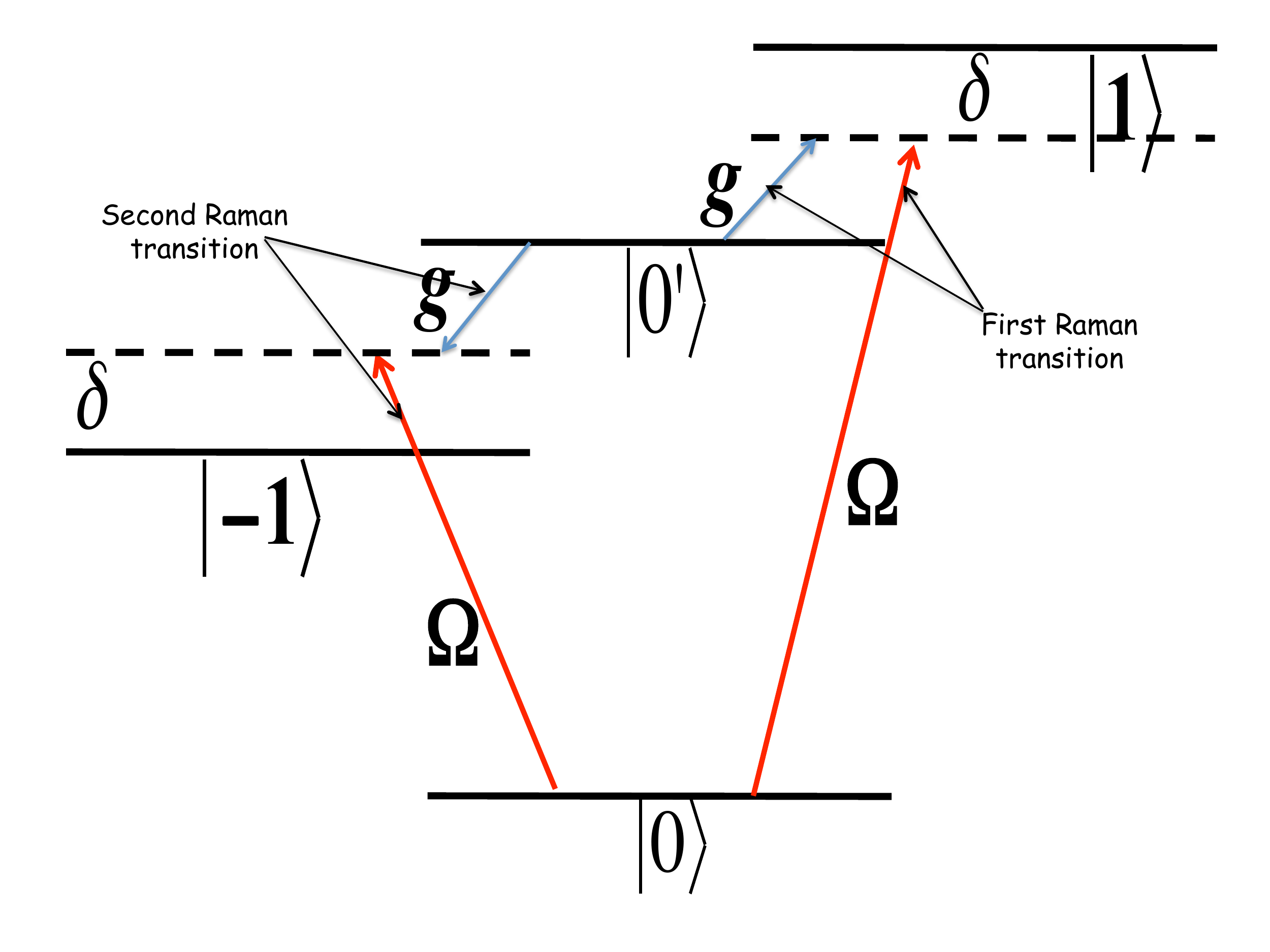}}
	\caption{Illustration of the frequency locking mechanism. An intuitive explanation of the frequency locking is evident in the bad limit($\delta \gg \Omega$) (in which the scheme is not efficient). The sensing signal  and the decoupling fields $\Omega$ stimulate a Raman transition between the two magnetic insensitive states. In this limit it is clear that the field that is being sensed is the one which is locked to the difference between the microwave decoupling fields.}
%and the Zeeman splitting. }
	\label{raman}
\end{figure}

Since the qubit's energy gap does not depend on the fluctuation of the magnetic field, the presented sensing mechanism is locked to the frequency difference between the microwave that couples $\vert -1 \rangle$ to $\vert 0 \rangle$ and the one that couples $\vert +1 \rangle$ to $\vert 0 \rangle$. 
%This is an important feature point of the new sensing method. 
The sensitivity of the method in the limit of large microwave drive depends on the relative frequency stability, i.e., on the fluctuations of the frequency difference between the two microwave drives. 
%in contrary to other methods that depend on the fluctuations in the Rabi frequency of the decoupling field. 
This can be easily seen in the opposite limit, when $\delta \gg \Omega$. In this limit (not useful for sensing) the states $\vert \pm 1 \rangle$ are only slightly populated and thus can be adiabatically eliminated. In this limit, state $\ket{B}$ is not populated and states $\ket u$ and $\ket d$ essentially are identical, up to phase factors,  with state $\ket{0}$). In this limit, we get two Raman transitions driven by the microwave fields and the signal field between the two magnetically insensitive states, $\ket {0'}$ and $\ket {0}$, one transition via the $\ket {+1}$ state and the other via the $\ket {-1}$ state (Fig. \ref{raman}).
Thus one sees that the locking of the signal is to the frequency difference between the microwave decoupling fields and not to the Rabi frequency of the microwave fields.

\end{document}